\documentclass[prb,reprint,twocolumn,showpacs,superscriptaddress,floatfix,aps,10pt]{revtex4-2}

\usepackage{amsmath, amsthm, amssymb}
\usepackage{dcolumn, bm, hyperref}
\usepackage{graphicx, subfigure, verbatim}
\usepackage{txfonts}    
\usepackage{newtxtext}
\usepackage{xcolor}
\usepackage{soul}
\usepackage{subfiles}
\usepackage{tabularx}

\renewcommand{\vec}[1]{{\bm{\mathrm{#1}}}}

\newcommand{\bra}[1]{\langle{#1}|}
\newcommand{\ket}[1]{|{#1}\rangle}

\usepackage{xcolor}

\begin{document}

\title{Magnetic octupole Hall effect in heavy transition metals}

\author{Insu Baek}
\affiliation{Department of Physics, Pohang University of Science and Technology, Pohang 37673, Korea}
\author{Seungyun Han}
\email{hanson@postech.ac.kr}
\affiliation{Department of Physics, Pohang University of Science and Technology, Pohang 37673, Korea}
\author{Hyun-Woo Lee}
\email{hwl@postech.ac.kr}
\affiliation{Department of Physics, Pohang University of Science and Technology, Pohang 37673, Korea}

\begin{abstract}
\textit{d}-wave altermagnets have the magnetic octupole as their primary order parameter. A recent study [Han \textit{et al.} arXiv 2409.14423 (2024)] demonstrated that magnetic octupole current can induce N\'eel vector dynamics. Therefore, identifying materials that can efficiently generate a magnetic octupole current is essential. In this paper, we investigate the magnetic octupole Hall effect in 4$d$ and 5$d$ transition metals. By employing atomic magnetic octupole operators, we calculate the magnetic octupole Hall conductivity using first-principles calculations. We also explore the microscopic origin of the magnetic octupole Hall effect and find that it results from the combined effect of orbital texture and spin-orbit coupling.
Additionally, we analyze the ratio of spin Hall conductivity to magnetic octupole Hall conductivity across various materials and identify those that are optimal for observing magnetic octupole physics. We also discuss potential applications arising from the magnetic octupole Hall effect. Our work serves as a valuable reference for identifying materials suitable for studying magnetic octupole physics.
\end{abstract}

\maketitle

\section{Introduction}

A spin current refers to the flow of spins, where spin-up electrons flow in one direction and spin-down electrons flow in opposite directions. When a spin current is injected into a magnetic material, it induces magnetization dynamics. This discovery has led to an extensive investigation into the interplay between local magnetization $\bold{M}$ and itinerant electron spin $\bold{S}$. For example, in ferromagnets (FMs), there exists $\bold{M} \cdot \bold{S}$ coupling. In equilibrium, the spins of itinerant electrons are aligned with the magnetization direction. When a spin current perpendicular to the magnetization direction is injected into FMs, the injected spins deviate from their equilibrium alignment, triggering magnetization dynamics and even switching the magnetization. Therefore, understanding how to generate a spin current and enhance its efficiency has become crucial, leading to extensive research. One of the primary methods for generating a spin current is the spin Hall effect (SHE), where a spin current flows in a direction perpendicular to the applied electric field. This approach enables the electrical generation of a spin current, which in turn allows for electrical control of magnetization, attracting significant attention and leading to numerous studies.

Recently, a new class of magnetic materials called altermagnets (AMs) has been proposed, which differs from the conventional classifications of FMs and antiferromagnets. Unlike these two well-known groups, AMs have no net spin magnetization and break Kramer spin degeneracy, nevertheless, resulting in \textit{d}-, \textit{g}-, or \textit{i}-wave spin splitting at $\bold{k}$ space $\bold{k}$ points~\cite{wu2007, hayami2019, naka2019, yuan2020, igor2021, yuan2021, ma2021, yuan2022_1, yuan2022_2}. There have been attempts to understand the characteristics of AMs in terms of an order parameter. In particular, in \textit{d}-wave AMs~\cite{bhowal2024}, a magnetic octupole (MO) is ordered. Furthermore, it was reported that there exists a linear coupling $\bold{N}\cdot \bold{O}_{ij}$ between the N\'eel vector $\bold{N}$ of the AM and the MO $\bold{O}_{ij}$, where $ij$ refers to the spatial index of MO \cite{mcclarty2024}.

Using an analogy to the spin current injection-induced magnetization dynamics in FM, injecting an MO current into an AM is expected to induce dynamics in the N\'eel vector. That is, the $\bold{N}\cdot \bold{O}_{ij}$ coupling, which tends to set the MO of an itinerant electron aligned along $\bold{N}$ in equilibrium, will induce the N\'eel vector dynamics when a MO current is injected with its spin polarization perpendicular to $\bold{N}$. Reference~\cite{han2024} addressed this issue and proposed two concepts: i) the MO current can be generated through the MO Hall effect (MOHE) in nonmagnets and ii) injection of a MO current into the AM generates torque to the system. This implies that similar to the magnetization dynamics achieved through the spin current injection into FM, the MO current injection serves as an effective tool for studying the N\'eel vector dynamics in AM. From this perspective, it is important to understand MOHE and identify materials where MOHE occurs significantly. However, Ref.~\cite{han2024} lacks a detailed investigation of its origin and the specific materials that exhibit significant MOHE. Reference~\cite{han2024} reported one material for MOHE: Pt. Since Pt also shows a strong spin Hall conductivity (SHC), it is crucial to find materials where the MO Hall conductivity (MOHC) is greater in comparison to SHC to study N\'eel vector dynamics induced dominantly by the MO current.

In this paper, we investigate the origin of MOHE and reveal that it is a combined effect of orbital texture and spin-orbit coupling (SOC). Furthermore, we systematically investigate MOHC in \textit{4d} and \textit{5d} transition metals. Considering the ubiquitous nature of orbital texture~\cite{go2018,han2023}, we expect that MOHE will occur not only in the \textit{4d} and \textit{5d} transition metals but also in materials with strong SOC. The paper is organized as follows. In Sec. II, we illustrate the concept of an atomic MO and review the atomic MO operators. In Sec. III, we introduce the MO Hall current as the MO version of the spin current and evaluate the MOHC by the first-principle calculation in \textit{4d} and \textit{5d} transition metal. In Sec. IV, we explain how the MO Hall current can be generated from the orbital texture and the SOC. In Sec. V, we compare the MOHCs in the \textit{4d} and \textit{5d} transition metals with their SHCs and propose the application of the MOHC. Our work will serve as a reference for selecting materials that can induce N\'eel vector dynamics by injecting MO into AM.

\section{Description of atomic magnetic octupole}

\subsection{Atomic MO picture}

\begin{figure}[t]
\includegraphics[width=8.5cm]{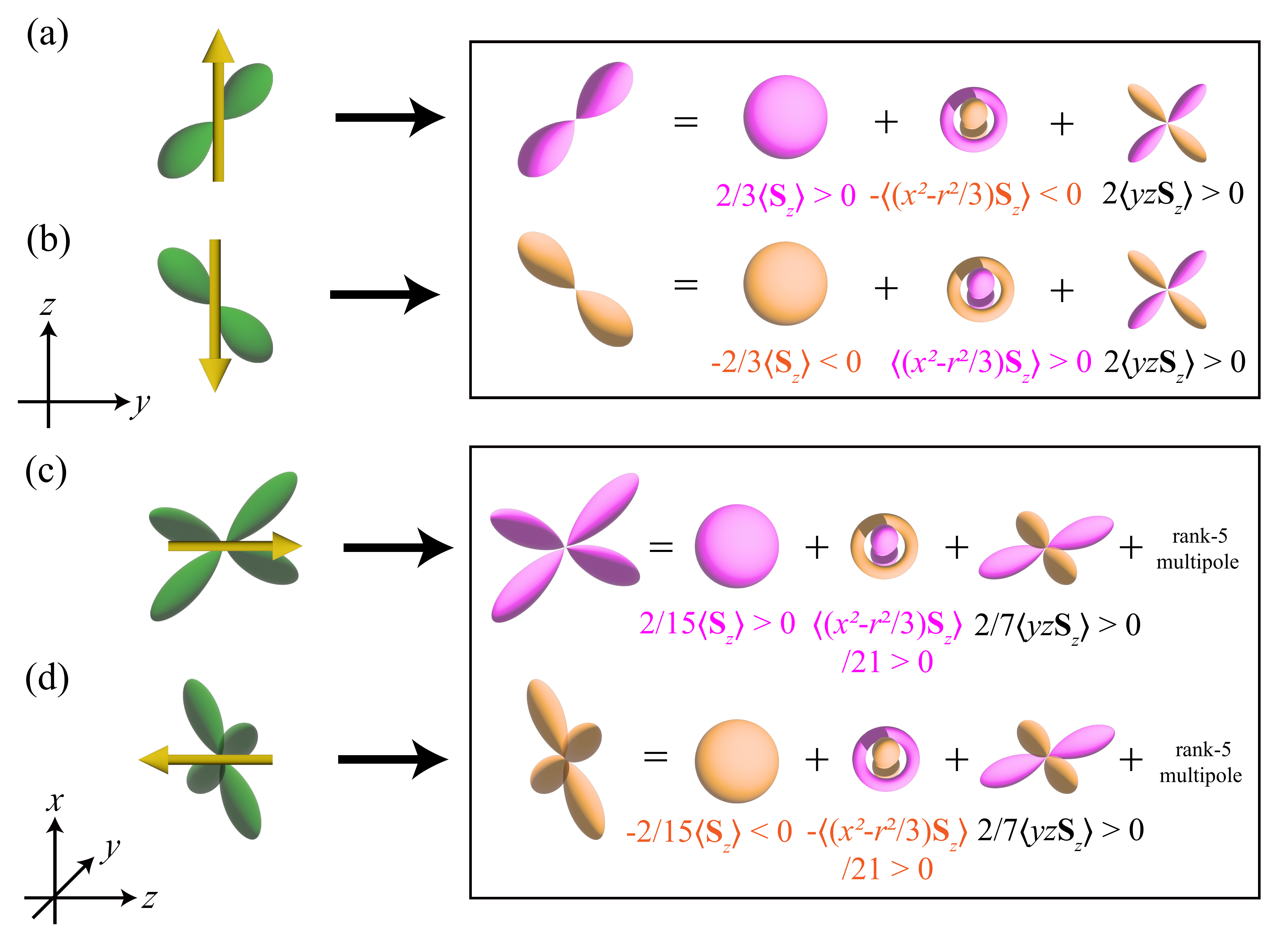}
 \caption{
Description of the atomic MO with positive $yzS_z$ density using real orbital states. The yellow arrows represent spin, and the greens represent electronic charge density. The pink (orange) color represents the positive (negative) spin density $S_z$. (a) For an electron with $\ket{p_y + p_z}$ orbitals carrying positive $S_z$ spin, the angular dependence of its spin density profile can be decomposed into a positive isotropic $S_z$ density, a negative $(x^2 - \frac{r^2}{3}) S_z$ density, and a positive $yzS_z$ MO density. (b) When an electron with $\ket{p_y - p_z}$ orbitals carrying negative $S_z$ spin, it can be decomposed into a negative isotropic $S_z$ density, a positive $(x^2 - \frac{r^2}{3}) S_z$ density, and a positive $yzS_z$ MO density. When (a) and (b) coexist in the equal contribution, the net spin density $\langle{S_z}\rangle$ and $\langle{(x^2 - \frac{r^2}{3})S_z}\rangle$ vanish, but the net MO density $\langle yzS_z\rangle$ remains finite. The spin density profile arising from \textit{d} electrons can be angular-decomposed in a similar way. (c) When an electron with $\ket{d_{xy} + d_{zx}}$ orbital carrying positive $S_z$ spin, the angular dependence of its spin density profile can be decomposed into a positive isotropic $S_z$ density, a positive $(x^2 - \frac{r^2}{3}) S_z$ density, a negative $yz S_z$ MO density, and higher-order magnetic multipole densities. (d) When an electron with $\ket{d_{xy} - d_{zx}}$ orbital carrying negative $S_z$ spin, the angular dependence of its spin density profile can be decomposed into a negative isotropic $S_z$ density, a negative $(x^2 - \frac{r^2}{3}) S_z$ density, a negative $yz S_z$ MO density, and higher-order magnetic multipole densities. When (c) and (d) coexist in the equal contribution, the net spin density $\langle{S_z}\rangle$ and $\langle{(x^2 - \frac{r^2}{3}) S_z}\rangle$ vanish, but the net MO density $\langle yzS_z\rangle$ remains finite. 
}
\label{fig:1}
\end{figure}

In this section, we explore how to describe an atomic MO. Atomic MO refers to a state in which the spin density distribution near an atomic site exists in the form of $r_n r_m S_q$, where $r_n$ represents a relative position with respect to the nearest atom centers and $S_m$ represents a spin operator. This form can be described using atomic orbitals and spins since atomic orbitals represent the spatial distribution of electrons. Therefore, the correlation between atomic orbitals and spins provides a framework for describing atomic MO. For example, if a $s$ orbital is spin-polarized, it results in an isotropic spin density. In contrast, when a $p$ orbital is spin-polarized, as shown in Fig.~\ref{fig:1} (a) and (b), the anisotropic spatial distribution of the $p$ orbital leads to a magnetic multipole density in the spin density profile. For the spin density profile of the state $\ket{p_y + p_z}$ in Fig.~\ref{fig:1}(a), the expectation value of $S_z(\theta, \phi)$ is given by $\langle S_z(\theta, \phi)\rangle \propto \sin^2(\theta) \cos^2(\phi - \pi/4) = \frac{1}{6} [4 + (3\cos^2(\theta) - 1) + 3 \sin^2(\theta) \cos(2\phi - \pi/2)]$, where $\tan \theta = \sqrt{y^2 + z^2}/x$ and $\tan \phi = z/y$. This can be decomposed into a positive isotropic spin-dipole density ($\langle S_z \rangle > 0$), a negative MO density $\langle (x^2 - \frac{r^2}{3}) S_z \rangle < 0$), and a positive MO density ($\langle yz S_z \rangle > 0$. For different $p$-orbital state $\ket{p_y - p_z}$ [Fig.~\ref{fig:1}(b)], we observe a negative isotropic spin density ($\langle S_z \rangle < 0$), a positive MO density ($\langle (x^2 - \frac{r^2}{3}) S_z \rangle < 0$), and a positive MO density ($\langle yz S_z \rangle > 0$). Due to symmetry constraints, both configurations are often equally populated. In such cases, the spin dipole densities cancel out, leaving a finite MO density. Thus, utilizing spin-polarized $p$ orbital states allows the description of atomic MO. Finite MO density can arise from $d$- and $f$-orbitals as well [Fig.~\ref{fig:1}(c)-(d)], although not only MO densities but also higher-order magnetic multipole densities simultaneously emerge in these cases, which is beyond the scope of this paper.

\subsection{Atomic MO operators}

Here, we show that the states described in Fig.~\ref{fig:1} can be captured using the operator $(1 / \hbar^2) \{L_m, L_n\} S_q$, where $L_m$ and $S_q$ are the atomic orbital angular momentum (OAM) and spin operators, respectively. The connection between $(1 / \hbar^2) \{L_m, L_n\} S_q$ and the MO can be illustrated as follows. For example, in the case of $(1 / \hbar^2) \{L_y, L_z\} S_z$, the eigenstates of $(1 / \hbar^2) \{L_y, L_z\}$ are $\ket{p_y \pm p_z} / \sqrt{2}$ and $\ket{p_x}$, with eigenvalues of $\mp 1$ and $0$, respectively. When combined with the spin operator, the eigenstates of $(1 / \hbar^2) \{L_y, L_z\} S_z$ with a negative eigenvalue are $\ket{p_y \pm p_z}\ket{s_z = \pm 1}/ \sqrt{2}$, corresponding to the states in Figs.~\ref{fig:1} (a) and (b), respectively. In fact, these operators are commonly used to capture atomic MO~\cite{jackeli2009, tahir2023, iwazaki2023}. This connection includes the other index choices $n, m. q$ of $(1 / \hbar^2) \{L_m, L_n\} S_q$. Thus, the atomic MO operator $O_{nm}^q$ that captures the MO density $r_n r_m S_q$ can be described by the following operator:
%%%%%%%%%%%%%%%%%%%%%%%%%%%%%%%
\begin{equation}\label{MOoperators}
    O_{nm}^q \equiv \frac{1}{\hbar^2}\{L_n,L_m\}S_q.
\end{equation}
%%%%%%%%%%%%%%%%%%%%%%%%%%%%%%%
In the following, we review the derivation of this relationship. Since the spin operators in Eq.~\eqref{MOoperators} capture the spin density in the MO, we demonstrate that $\langle 1 / \hbar^2 \{L_n, L_m\} \rangle \sim \langle r_n r_m \rangle$ around the atomic site. 

With the spherical harmonics, atomic electric multipole operator $\hat{Q}_{lm}^{\rm{orb}}$ is defined as~\cite{kusunose2020}
\begin{subequations}\label{eq:1}
    \begin{align}
        \bra{n_1 l_1 m_1} \hat{Q}_{lm}^{\rm{orb}} \ket{n_2 l_2 m_2} &= \bra{l_1 m_1} C_{lm} \ket{l_2 m_2} \bra{n_1 l_1} r^l \ket{n_2 l_2}, \label{eq:1b} \\ 
        \bra{n_1 l_1} r^l \ket{n_2 l_2} &=  
        \int dr \ r^{l + 2} R_{n_1 l_1} (r) R_{n_2 l_2} (r), \label{eq:1c} \\
        \bra{l_1 m_1} C_{lm} \ket{l_2 m_2} &=
        (-1)^{m_1} \sqrt{(2l_1 + 1) (2l_2 + 1)} \nonumber \\
        & \times
        \begin{pmatrix}
            l_1 & l_2 & l \\
            - m_1 & m_2 & m
        \end{pmatrix}
        \begin{pmatrix}
            l_1 & l_2 & l \\
            0 & 0 & 0
        \end{pmatrix}, \label{eq:1d}
    \end{align}
\end{subequations}
where \textit{n}, \textit{l}, \textit{m} are the principal, azimuthal, magnetic quantum numbers, respectively, $ R_{nl}(\vec{r})$ is the radial function and the spherical harmonics, and $(\begin{smallmatrix} l_1 & l_2 & l \\ - m_1 & m_2 & m \end{smallmatrix})$ is the Wigner's \textit{3j} symbol. For $l = 2$ corresponds to an atomic electric quadrupole $\propto r_nr_m$. The orbitals with nonzero $l_1$ and $l_2$, such as \textit{p}, \textit{d}, and \textit{f} orbitals can generate the atomic electric quadrupole $\hat{Q}_{2m}^{\rm{orb}}$. Here, we focus only on \textit{p} and \textit{d} orbitals since we deal with transition metals. $\bra{n_1 l_1} r^l \ket{n_2 l_2}$ and $\bra{l_1 m_1} C_{lm} \ket{l_2 m_2}$ are the operators which represent the radial $r^l$ and angular $C_{lm} \propto Y_{lm}$ components, which is the spherical harmonics, between two atomic-centered states $\ket{n_1 l_1 m_1}$ and $\ket{n_2 l_2 m_2}$, respectively. Since $n_1 = n_2$ and $l_1 = l_2 = l$, we drop the indices in $\bra{n_1 l_1} r^l \ket{n_2 l_2}$ and $\bra{l_1 m_1} C_{lm} \ket{l_2 m_2}$, and just keep $\langle r^l \rangle$ and $C_{lm}$, respectively. 
For \textit{p}-orbitals ($n_1 = n_2 = n, \ l_1 = l_2 = 1$), the electric quadrupole operator $\bra{n_1 l_1 m_1} \hat{Q}_{2m}^{\rm{orb}} \ket{n_2 l_2 m_2}$ using the cubic harmonic basis $\Big(\ket{p_x}, \ket{p_y}, \ket{p_z} \Big)$~\cite{Kwiatkowski1978} is reformed as~\cite{hayami2024}
\begin{subequations}\label{eq:3}
    \begin{align}
        x^2 - y^2 &= \langle r^2 \rangle \frac{C_{2 -2} + C_{2 2}}{\sqrt{2}} = - \frac{\sqrt{3}}{5} \frac{1}{\hbar^2} \langle r^2 \rangle \Big( L_x^2 - L_y^2 \Big),
        \label{eq:3a} \\
        zx &= \langle r^2 \rangle \frac{C_{2 -1} - C_{2 1}}{\sqrt{2}} = - \frac{\sqrt{3}}{5} \frac{1}{\hbar^2} \langle r^2 \rangle \Big\{L_x, L_z \Big\},
        \label{eq:3b} \\
        3z^2 - r^2 &= \langle r^2 \rangle C_{2 0} = - \frac{\sqrt{3}}{5} \frac{1}{\hbar^2}\langle r^2 \rangle\frac{2L_z^2 - L_x^2 - L_y^2}{\sqrt{3}},
        \label{eq:3c} \\
        yz &= \langle r^2 \rangle \frac{i(C_{2 -1} + C_{2 1})}{\sqrt{2}} = - \frac{\sqrt{3}}{5} \frac{1}{\hbar^2}\langle r^2 \rangle\Big\{L_y, L_z \Big\},
        \label{eq:3d} \\
        xy &= \langle r^2 \rangle \frac{i(C_{2 -2} - C_{2 2})}{\sqrt{2}} = - \frac{\sqrt{3}}{5} \frac{1}{\hbar^2}\langle r^2 \rangle\Big\{L_x, L_y \Big\}.
        \label{eq:3e}
    \end{align}
\end{subequations}
Note that the product of two position operators can be expressed in terms of the anticommutator of the atomic OAM operators ($\{L_i, L_j\} / \hbar^2 \sim r_i r_j$) since the radial and the angular parts of the Bloch wavefunctions are separated into $\langle r^l \rangle$ and $ C_{lm}$ respectively. Combined with spin operators, we can confirm from this that $(1 / \hbar^2) \{L_m, L_n\} S_q$ in Eq.~\eqref{MOoperators} captures the atomic MO. To accurately calculate the atomic MO density, one must use Eq.~\eqref{eq:1} for determining proportional factor, e.g., $-\sqrt{5}\langle r^2 \rangle/3$ for $p$-orbitals. However, as previously defined in Eq.~\eqref{MOoperators}, we utilize an operator where $-\sqrt{5}\langle r^2 \rangle/3$ is set to unity. This definition makes the quantity dimensionally equivalent to spin and OAM, allowing for easier comparison of their relative magnitudes. To convert this into MO density units, one can apply the proportionality constant defined in Eq.~\eqref{eq:3} and multiply by $-\sqrt{5}\langle r^2 \rangle/3$. Similar conclusions can be drawn for \textit{d} orbitals ($n_1 = n_2 = n, \ l_1 = l_2 = 2$) where the proportional constant in Eq.~\eqref{eq:3} is replaced by $-\sqrt{3}/21$ and the $L_i$ operators for \textit{p} orbitals are replaced by $L_i$ for \textit{d} orbitals~\cite{jo2018}. As a side remark, $\{L_i, L_j\} / \hbar^2$ corresponds to the orbital angular position (OAP) operators introduced in Ref.~\cite{han2022}, which describe the angular position of the orbital state.
The dynamics of the OAP are intertwined with the dynamics of the OAM and play a fundamental role in OAM dynamics. From this, we can expect that the dynamics of MO, which is captured by OAP times spin operators, could naturally emerge in materials where traditional OAM dynamics are investigated.  

\section{Magnetic octupole Hall effect}

\subsection{MO current}

\begin{figure}[t]
\includegraphics[width=8.5cm]{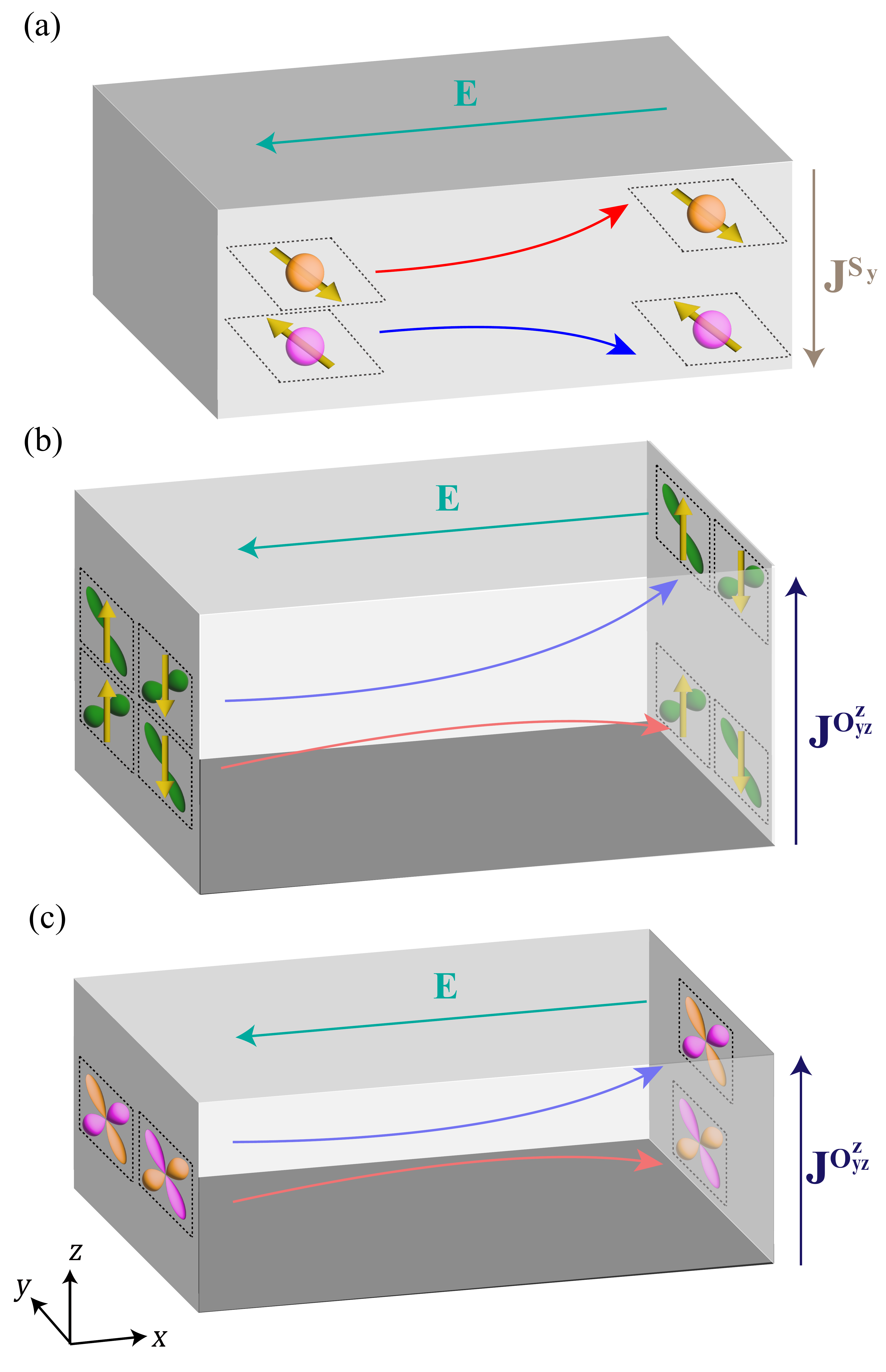}
 \caption{
Description of MOHE compared to SHE. (a) When an electric field $\bold{E}$ induces an electron with spin-up to flow upward while that with spin-down to flow in the opposite direction, this corresponds to the SHE. (b) When an electric field induces an orbital correlated with spin to flow upward (i.e., $\ket{p_y \pm p_z}\ket{s_z \pm}$) and the oppositely correlated state (i.e., $\ket{p_y \pm p_z}\ket{s_z \mp}$) to flow downward, it corresponds to the MOHE. (c) The spin-polarized orbitals $\ket{p_y \pm p_z}\ket{s_z \pm}$ have zero spin density and nonzero MO density $yzS_z$ [Fig.~\ref{fig:1} (a) and (b)] so that the MO current in (b) can be represented by the MO density $\pm yzS_z$ flowing along $\pm z$ directions.
 }
\label{fig:5}
\end{figure}

In addition to the MO density, it is possible to define the MO current as one does the spin current. A spin current is defined by $J_j^{S_k} = 1/2 \{v_j, S_k\}$, where $v_j$ is the $j$th component of the velocity operator. Figure~\ref{fig:5}(a) shows the spin current $J_z^{S_y}$, which is spin-polarized along $S_y$ flowing in the $-\hat{z}$ direction. Electrons with spin $S_y > 0$ ($S_y < 0$) flow along the direction -$\hat{z}$ ($\hat{z}$), so the spin $S_y$ current flows along the $z$ direction without charge current~\cite{murakami2003, sinova2004}. Likewise, using the definition of the MO operator, the MO current can be defined as $J_j^{O^q_{mn}} = 1/2 \{v_j, O^q_{mn}\}$. Figure~\ref{fig:5}(b) illustrates the MO current $J_z^{O_{yz}^z}$ with the MO component $O_{yz}^z$ flowing in the $\hat{z}$ direction. Electrons with $\ket{p_y + p_z} (\ket{p_y - p_z})$ orbitals polarized with spin $\ket{s_z+} (\ket{s_z-})$, which have positive MO density ($\propto yzS_z > 0$) flow along the $z$ while electrons with $\ket{p_y + p_z}(\ket{p_y - p_z})$ orbitals polarized with spin $\ket{s_z-} (\ket{s_z+})$, which have negative MO density ($\propto yzS_z < 0$) flow along the $-z$ [Fig.~\ref{fig:5}(c)]. Note that in Fig.~\ref{fig:5}(b), there is no net charge or spin current; the MO current with the $O^y_{yz}$ component flows along the $z$ direction.

\subsection{First-principle calculation of the MO Hall effect}

For electrical generation of an MO current, we consider the MOHE, where a MO current flows perpendicular to an electric field ($\vec{\mathcal{E}}$). This relation may be summarized as follows,
\begin{equation}\label{MOHE}
    J_j^{O^q_{mn}}= \chi_{ji}^{O_{mn}^q} \mathcal{E}_i,
\end{equation}
where $i$ and $j$ denote orthogonal directions and $\chi_{ji}^{O_{mn}^q}$ amounts to the MOHC. First, we investigate the symmetrically allowed Hall components of MOHE. For this, we consider centrosymmetric time-reversal systems with mirror symmetries $M_x$ and $M_y$ with an electric field applied along the $x$ direction and investigate the $z$ flow of a MO current. Under these constraints, the allowed MO current components are $J_z^{O_{xy}^x}$, $J_z^{O_{yz}^z}$, $J_z^{O_{xx}^y}$, $J_z^{O_{yy}^y}$, and $J_z^{O_{zz}^y}$, where the second component is illustrated in Fig.~\ref{fig:5}(b). Under the same symmetry constraints, the only symmetrically allowed spin current is $j_z^{S_y}$ [Fig.~\ref{fig:5}(a)], whereas $j_z^{S_x}$ and $j_z^{S_z}$ are forbidden. Therefore, when decomposing the electrons flowing along the $z$-axis into magnetic multipoles, we find that for the spin polarization along in the $x$- and $z$-directions, the MO is the leading term. This is because the spin Hall current cannot be spin-polarized along the $x$- and $z$-directions. Thus, we focus our investigation on these two components, $J_z^{O_{xy}^x}$ and $J_z^{O_{yz}^z}$.

To evaluate the MOHC, we employ the Kubo formula within the linear response theory using first-principle calculations. The MOHC $\chi_{ji}^{O_{mn}^{q}}$ is given by
\begin{align}\label{eq:17}
    \chi_{ji}^{O_{mn}^{q}} & = \frac{e}{\hbar} \sum\limits_{\mu \neq \nu} \int \frac{d^3 k}{(2 \pi)^3}  \\
&\quad\times (f_{\mu\textbf{k}}-f_{\nu\textbf{k}}) \hbar^2 \operatorname{Im}\Big[\frac{\bra{u_{\mu\textbf{k}}}{\frac{1}{2} \{v_j, O_{mn}^q\}}\ket{u_{\nu\textbf{k}}}\bra{u_{\nu\textbf{k}}}{v_i}\ket{u_{\mu\textbf{k}}}}{(E_{\mu\textbf{k}}-E_{\nu\textbf{k}})(E_{\mu\textbf{k}}-E_{\nu\textbf{k}} + i\Gamma)}\Big]\nonumber,
\end{align}
where $e$ is the electronic charge and we set $\Gamma = 0.0259 \ \rm{eV}$ which is the energy level broadening at room temperature, $f_{\mu\textbf{k}}$ is the Fermi-Dirac distribution function, $\ket{u_{\mu\textbf{k}}}$ is a periodic part of the Bloch state with the energy eigenvalue $E_{\mu\textbf{k}}$. The temperature is set to $T = 300 \ \rm{K}$. Note that the dimension of $\chi_{ji}^{O_{mn}^{l}}$ is $(\hbar/e) \ (\Omega \ \mathrm{cm})^{-1}$, which is the same unit of the SHC, since we set the dimension of MO to be the same as that of spin. This allows us to directly compare the magnitude of the MOHC with that of the SHC.
%%%
\begin{figure}[t]
\includegraphics[width=8.5cm]{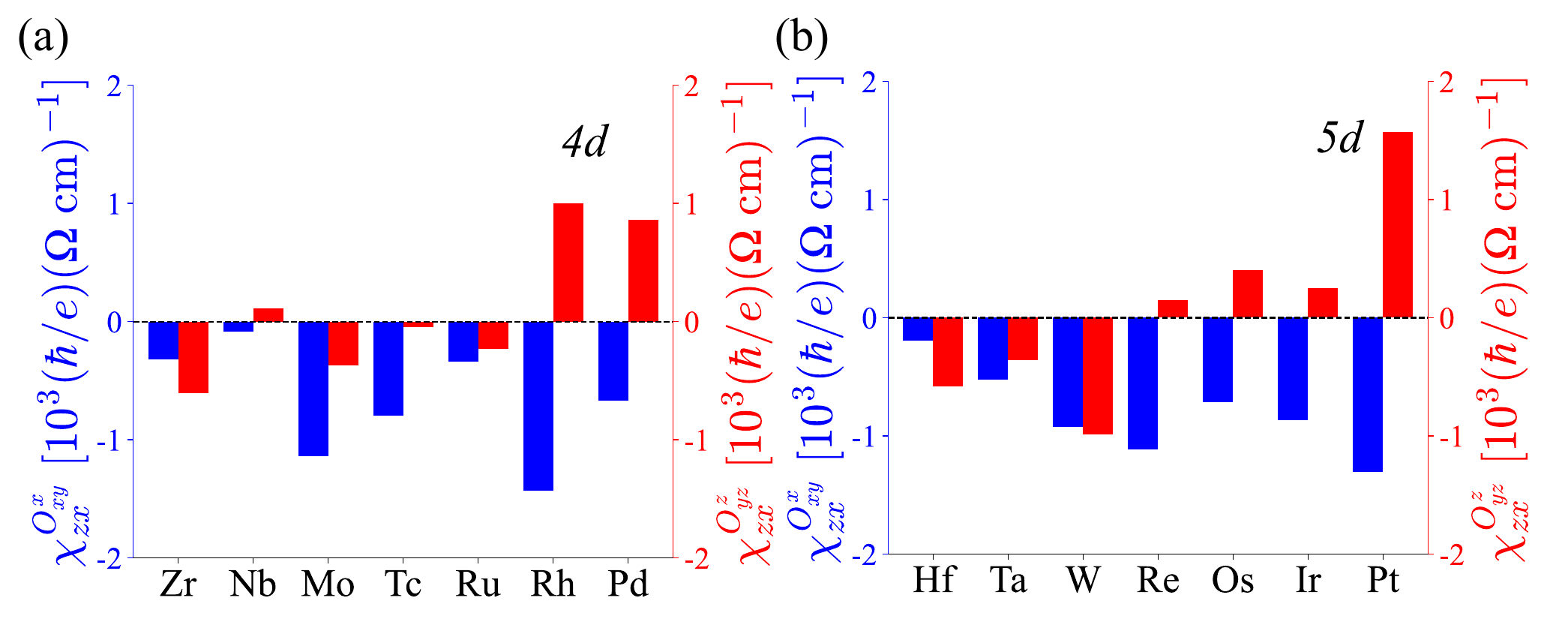}
 \caption{
Magnetic octupole Hall conductivity $\chi_{zx}^{O_{xy}^{x}}$ (blue) and $\chi_{zx}^{O_{yz}^{z}}$ (red) for \textit{4d} (a) and \textit{5d} (b) transition metals at the Fermi energy with unit $(\hbar/e) \ (\Omega \ \rm{cm})^{-1}$.
 }
\label{fig:3}
\end{figure}
%%%
We calculate the MOHC for \textit{4d} (hcp Zr, bcc Nb, bcc Mo, hcp Tc, hcp Ru, fcc Rh, and fcc Pd) and \textit{5d} transition metals (hcp Hf, bcc Ta, bcc $\alpha$-W, hcp Re, hcp Os, fcc Ir, and fcc Pt) using the full-potential DFT calculation as follows. First, we obtain self-consistent electronic structures using the full potential linearization augmented wave method~\cite{FLAPW} with the FLEUR code~\cite{FLEUR}. The Perdew-Burke-Ernzerhof exchange-correlation functional is used within the generalized gradient approximation~\cite{PBE}. The Brillouin zone is sampled using the $16 \times 16 \times 16$ Monkhorst-Pack \textbf{k}-point mesh~\cite{MPgrid}. We take the lattice constant, the muffin-tin radius, and the plane wave cutoffs for each material from~\cite{go2024}.
%%%
Next, we obtain the maximally localized Wannier functions (MLWFs) from the Bloch states with the WANNIER90 code \cite{WANNIER90}. The Brillouin zones are sampled with the equidistant $8 \times 8 \times 8$ \textbf{k}-mesh, which includes the $\Gamma$ point. The Bloch states are initially projected into the $s, \ p_x, \ p_y, \ p_z, \ d_{xy}, \ d_{yz}, \ d_{xz}, \ d_{x^2 - y^2}$ and $d_{z^2}$ states. The 18 MLWFs for each atomic site are chosen out of 36 bands. The frozen energy windows are set to include a region of 5 eV higher than the Fermi energy. 
We evaluate $v_{i}$ by including the anomalous position~\cite{go2024}. The \textbf{k} integration is calculated using a uniformly distributed 150 $\times$ 150 $\times$ 150 \textbf{k}-mesh grid.

\begin{table}
\centering
\caption{
Magnetic octupole Hall conductivity $\chi_{zx}^{O_{xy}^{x}}$, $\chi_{zx}^{O_{yz}^{z}}$ and spin Hall conductivity $\sigma_{zx}^{y}$~\cite{go2024} for \textit{4d} and \textit{5d} transition metals at the Fermi energy with unit $(\hbar/e) \ (\Omega \ \rm{cm})^{-1}$. The lattice vectors of hcp materials are defined as $\bold{a}_1 = a (\frac{\sqrt{3}}{2}, -\frac{1}{2}, 0)$, $\bold{a}_2 = a (\frac{\sqrt{3}}{2}, \frac{1}{2}, 0)$, and $\bold{a}_3 = c (0, 0, 1)$ for lattice constants $a$ and $c$, respectively.
}
\begin{tabularx}{0.45\textwidth} { 
   >{\raggedright\arraybackslash}X 
   >{\centering\arraybackslash}X
   >{\centering\arraybackslash}X
   >{\centering\arraybackslash}X  }
 \hline
 \hline
 Materials & $\chi_{zx}^{O_{xy}^{x}}$ & $\chi_{zx}^{O_{yz}^{z}}$ & $\sigma_{zx}^{y}$ \\ 
 \hline
 hcp Zr &   -319    &   -607  &  -30  \\
 bcc Nb &    -82    &    111  &  -74  \\
 bcc Mo &   -1138   &   -370  &  -254  \\
 hcp Tc &   -794    &    -49  &  -72  \\
 hcp Ru &   -340    &   -232 &   135  \\
 fcc Rh &   -1432   &   1000 &   987  \\
 fcc Pd &   -668    &    862  &  1111 \\
 hcp Hf &   -194    &   -584 &   50   \\
 bcc Ta &   -523    &   -356  &  -160 \\
 bcc W  &   -926    &   -990  &  -788 \\
 hcp Re &   -1117   &    149  &  -456 \\
 hcp Os &   -717    &    403  &  -40  \\
 fcc Ir &   -864    &    248  &  321  \\
 fcc Pt &   -1303   &   1569  &  2212 \\
 \hline
 \hline
\end{tabularx}
\label{tab:1}
\end{table}

Figure~\ref{fig:3} shows the calculated MOHCs $\chi_{zx}^{O_{xy}^{x}}$ (blue line) and $\chi_{zx}^{O_{yz}^{z}}$ (red line) for \textit{4d} [Fig.~\ref{fig:3}(a)] and \textit{5d} transition metals [Fig.~\ref{fig:3}(b)] at Fermi energy. Their numerical values are listed in Table~\ref{tab:1}, compared to the SHCs from Ref~\cite{go2024}. Transition metals have large MOHC $\sim 10^2 \ (\hbar/e) \ (\Omega \ \rm{cm})^{-1}$, and some of them, such as bcc Mo, fcc Rh, bcc W, hcp Re, and fcc Pt, have gigantic MOHCs $\sim 10^3 \ (\hbar/e) \ (\Omega \ \rm{cm})^{-1}$. The maximum intensities of MOHCs are $\chi_{zx}^{O_{xy}^{x}} = -1432 \ (\hbar/e) \ (\Omega \ \rm{cm})^{-1}$ (fcc Rh) and $\chi_{zx}^{O_{yz}^{z}} = 1569 \ (\hbar/e) \ (\Omega \ \rm{cm})^{-1}$ (fcc Pt). They originate from the strong SOC of fcc Rh and Pt. The $\chi_{zx}^{O_{xy}^{x}}$ components are negative in every transition metal. For the $\chi_{zx}^{O_{yz}^{z}}$ component, transition metals with $\langle \vec{L} \cdot \vec{S} \rangle < 0$ such as hcp Zr, bcc Nb, bcc Mo, hcp Hf, bcc Ta and bcc W have negative signs, while transition metals with $\langle \vec{L} \cdot \vec{S} \rangle > 0$ such as fcc Rh, fcc Pd, fcc Rh, and fcc Pt have positive signs. The sign dependence of $\chi_{zx}^{O_{yz}^{z}}$ on $\langle \vec{L} \cdot \vec{S} \rangle$ is analogous to the sign dependence of the SHC in transition metals~\cite{tanaka2008, kontani2009, jo2018, salemi2022}.

\section{Microscopic origin of MOHE}

\begin{figure}[t]
\includegraphics[width=8.5cm]{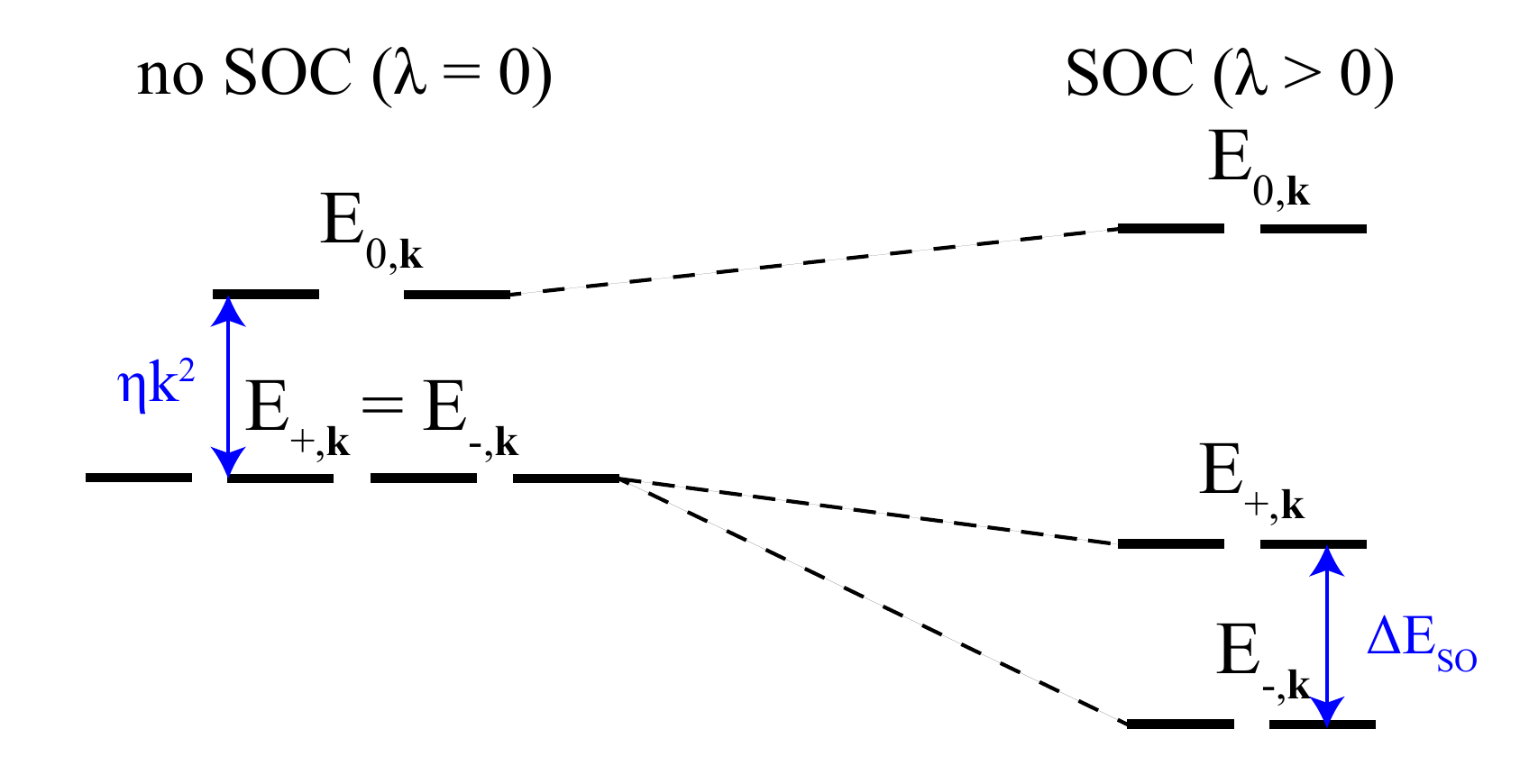}
 \caption{ 
The eigenenergies of the Hamiltonian $\mathcal{H}$ with no SOC ($\lambda = 0$) and with SOC ($\lambda > 0$). When the SOC $\lambda$ is turned on, the energy $E_{+,\bold{k}}$ and $E_{-,\bold{k}}$ are splitted by $\Delta E_{\rm{SO}}$.
 }
\label{fig:2}
\end{figure}

As demonstrated in the previous section, MOHE is ubiquitous. This is because MOHE shares the same underlying origin as the spin Hall effect—namely, the orbital Hall effect (OHE) combined with SOC~\cite{go2018,han2024}—a mechanism widely observed in multiorbital systems. In further detail, a multiorbital system generally possesses an OAP texture~\cite{han2023}, which is the origin of the OHE~\cite{go2018}. When a perturbation is applied, such as an electric field, the OAP dynamics is induced, leading to the occurrence of OHE~\cite{go2018,han2022}. Up to this point, the phenomenon is independent of SOC. When SOC is introduced, the OAP dynamics becomes intertwined with the MO dynamics, resulting in the MOHE. Although MOHE is discussed in Ref.~\cite{han2024}, it does not provide a detailed microscopic picture of this process. Here, we will illustrate this through a low-energy Hamiltonian.

\subsection{Model Hamiltonian}
For this, we adopt a Hamiltonian with a $p$ orbital texture~\cite{han2023} and the SOC strength $\lambda$. We take the following low-energy Hamiltonian $\mathcal{H}$:
\begin{equation}\label{eq:7}
 \mathcal{H} = \frac{\hbar^2 k^2}{2m} -\eta \left(\textbf{L}\cdot \textbf{k}\right)^2  + \lambda \vec{L} \cdot \vec{S},
\end{equation}
where \textit{m} is the electron mass, \vec{k} is the crystal momentum, $\eta$ is the crystal field strength for the orbital texture, and \vec{L} and \vec{S} are the \textit{p}-orbital OAM and the spin operators. 

\subsection{OHE}
We first review the microscopic origin of the OHE, which occurs even when $\lambda=0$~\cite{go2018}. In this case, the spin degree of freedom is decoupled from the orbital dynamics, allowing us to focus solely on the orbital degrees of freedom. This results in energy splitting between a radial state denoted as $\ket{p_{\bold{k}}}$, whose $p$ orbital lobe is aligned with the vector $\bold{k}$, and two degenerate tangential states, $\ket{p_{\theta_{\bold{k}}}}$ and $\ket{p_{\phi_{\bold{k}}}}$, where $\theta_{\bold{k}}$ and $\phi_{\bold{k}}$ are defined through the relation $\bold{k} = (k_x, k_y, k_z) = |\bold{k}|(\sin{\theta_{\bold{k}} \cos{\phi_{\bold{k}}}}, \cos{\theta_{\bold{k}}}, \sin{\theta_{\bold{k}} \sin{\phi_{\bold{k}}}})$. We investigate the generation of the orbital current $j_z^{L_y}$ when an electric field is applied along the $x$ direction, $\mathcal{E}_x$. For simplicity of illustration, we focus on the $k_y=0$ plane, although the illustrated process can be applied to the entire $\bold{k}$ space. In this case, the radial orbital and two tangential orbitals become $\ket{p_{\bold{k}}}=\cos\phi_\bold{k} \ket{p_x}+\sin\phi_\bold{k}\ket{p_z}$, $\ket{p_{\phi_{\bold{k}}}}=-\sin\phi_\bold{k} \ket{p_x}+\cos\phi_\bold{k}\ket{p_z}$, and $\ket{p_{\theta_\bold{k}}}=\ket{p_y}$ [Fig.~\ref{fig:6}(a)]. Note that the $\ket{p_y}$ orbital is not mixed with other orbitals in the $k_y=0$ plane. When an electric field is applied along $x$, the crystal momentum $\bold{k}$ is shifted to $\bold{k}+\delta\bold{k}$, where $\delta\bold{k}$ is along the $-x$ direction. Then $\ket{p_{\phi_{\bold{k}}}}$ can be decomposed into
\begin{align}
    \ket{p_{\phi_{\bold{k}}}} = \ket{p_{\phi_{\bold{k}+\delta\bold{k}}}}+\delta\phi\ket{p_{\bold{k}+\delta\bold{k}}},
\end{align}
where $\delta \phi \propto \mathcal{E}_x$. Then, after a short time $\delta t$, the two terms acquire different phase factors, $\ket{p_{\phi_{\bold{k}}},\delta t} = \ket{p_{\phi_{\bold{k}+\delta\bold{k}}}}e^{-iE_{\phi_{\bold{k}+\delta\bold{k}}}\delta t}+\delta\phi\ket{p_{\bold{k}+\delta\bold{k}}}e^{-iE_{\bold{k}+\delta\bold{k}}\delta t}$, where $E_{\phi_{\bold{k}+\delta\bold{k}}}$ and $E_{\bold{k}+\delta\bold{k}}$ are the energies of $\ket{p_{\phi_{\bold{k}+\delta\bold{k}}}}$ and $\ket{p_{\bold{k}+\delta\bold{k}}}$, respectively. For this state, we calculate the expectation value of $L_y$ operator, which results in $\delta\phi_\bold{k}\text{Im}[e^{-i(E_{\bold{k}+\delta\bold{k}}-E_{\phi_{\bold{k}+\delta\bold{k}}})\delta t}] = -\delta\phi_\bold{k} (E_{\bold{k}+\delta\bold{k}}-E_{\phi_{\bold{k}+\delta\bold{k}}})\delta t$. Straightforward calculation shows that $\delta\phi_\bold{k}\propto k_z$, which results in an OAM current along $z$ direction, $j_z^{L_y}$  [Fig.~\ref{fig:6}(b)]. In summary, mixing between the radial and the tangential orbital leads to OHE.

\subsection{MOHE}

Now, we turn on SOC $\lambda$ and illustrate the origin of MOHE. Specifically, we illustrate the flow in the $z$ direction of MO currents $j_z^{O_{xy}^x}$ and $j_z^{O_{yz}^z}$ induced by an electric field $\mathcal{E}_x$ applied along the $x$ direction. In this case, due to SOC, the eigenstates deviate from the previous case. Specifically, the previously degenerate two tangential bands in Sec. IV B (four tangential bands, when the spin degree of freedom is considered) undergo spin-dependent splitting (although, due to Kramer's degeneracy, there is no net spin polarization in each band). The eigenstates are given by

\begin{align}
    \ket{\psi_{1,\bold{k}}} &= \cos\alpha \ket{p_\bold{k}}\ket{\uparrow_\bold{k}}+\sin\alpha \ket{p_{+,\bold{k}}}\ket{\downarrow_\bold{k}} \\
    \ket{\psi_{2,\bold{k}}} &= \cos\alpha \ket{p_\bold{k}}\ket{\downarrow_\bold{k}}-\sin\alpha \ket{p_{-,\bold{k}}}\ket{\uparrow_\bold{k}} \\
    \ket{\psi_{3,\bold{k}}} &= -\sin\alpha \ket{p_\bold{k}}\ket{\uparrow_\bold{k}}+\cos\alpha \ket{p_{+,\bold{k}}}\ket{\downarrow_\bold{k}} \\
    \ket{\psi_{4,\bold{k}}} &= \sin\alpha \ket{p_\bold{k}}\ket{\downarrow_\bold{k}}+\cos\alpha \ket{p_{-,\bold{k}}}\ket{\uparrow_\bold{k}} \\
    \ket{\psi_{5,\bold{k}}} &= \ket{p_{+,\bold{k}}}\ket{\uparrow_\bold{k}} \\
    \ket{\psi_{6,\bold{k}}} &= \ket{p_{-,\bold{k}}}\ket{\downarrow_\bold{k}},
\end{align}
where $\bold{S} \cdot \hat{\bold{k}} \ket{\uparrow_\bold{k}}=\hbar/2 \ket{\uparrow_\bold{k}}$, $\bold{S}\cdot\hat{\bold{k}}\ket{\downarrow_\bold{k}}=-\hbar/2 \ket{\downarrow_\bold{k}}$, $2\alpha = \tan^{-1} \Big[2\sqrt{2} \lambda/(\eta k^2 +\lambda \Big)]$, and $\ket{p_{\pm,\bold{k}}}=(\ket{p_{\phi_\bold{k}}}\pm i\ket{p_{\theta\bold{k}}})/\sqrt{2}$. Using these states, we can diagonalize $\mathcal{H}$ with the following eigenenergies [Fig.~\ref{fig:2}(a)]
\begin{subequations}\label{eq:10}
    \begin{align}
        E_{1,\bold{k}}&= E_{2,\bold{k}} = E_{0,\bold{k}} \nonumber \\
        &= \frac{\hbar^2 k^2}{2m} - \frac{(\eta k^2 +\lambda)}{2} + \sqrt{\left(\frac{\eta k^2 +\lambda}{2}\right)^2 +2\lambda^2}, \label{eq:10c} \\
        E_{3,\bold{k}} &= E_{4,\bold{k}} = E_{-,\bold{k}}  \nonumber \\
        &= \frac{\hbar^2 k^2}{2m}- \frac{(\eta k^2 +\lambda)}{2} - \sqrt{\left(\frac{\eta k^2 +\lambda}{2}\right)^2 +2\lambda^2}, \label{eq:10a} \\
        E_{5,\bold{k}}&=E_{6,\bold{k}} = E_{+,\bold{k}} = \frac{\hbar^2 k^2}{2m}-\eta k^2 +\lambda. \label{eq:10b}
    \end{align}
\end{subequations}

\begin{figure}[t]
\includegraphics[width=8.5cm]{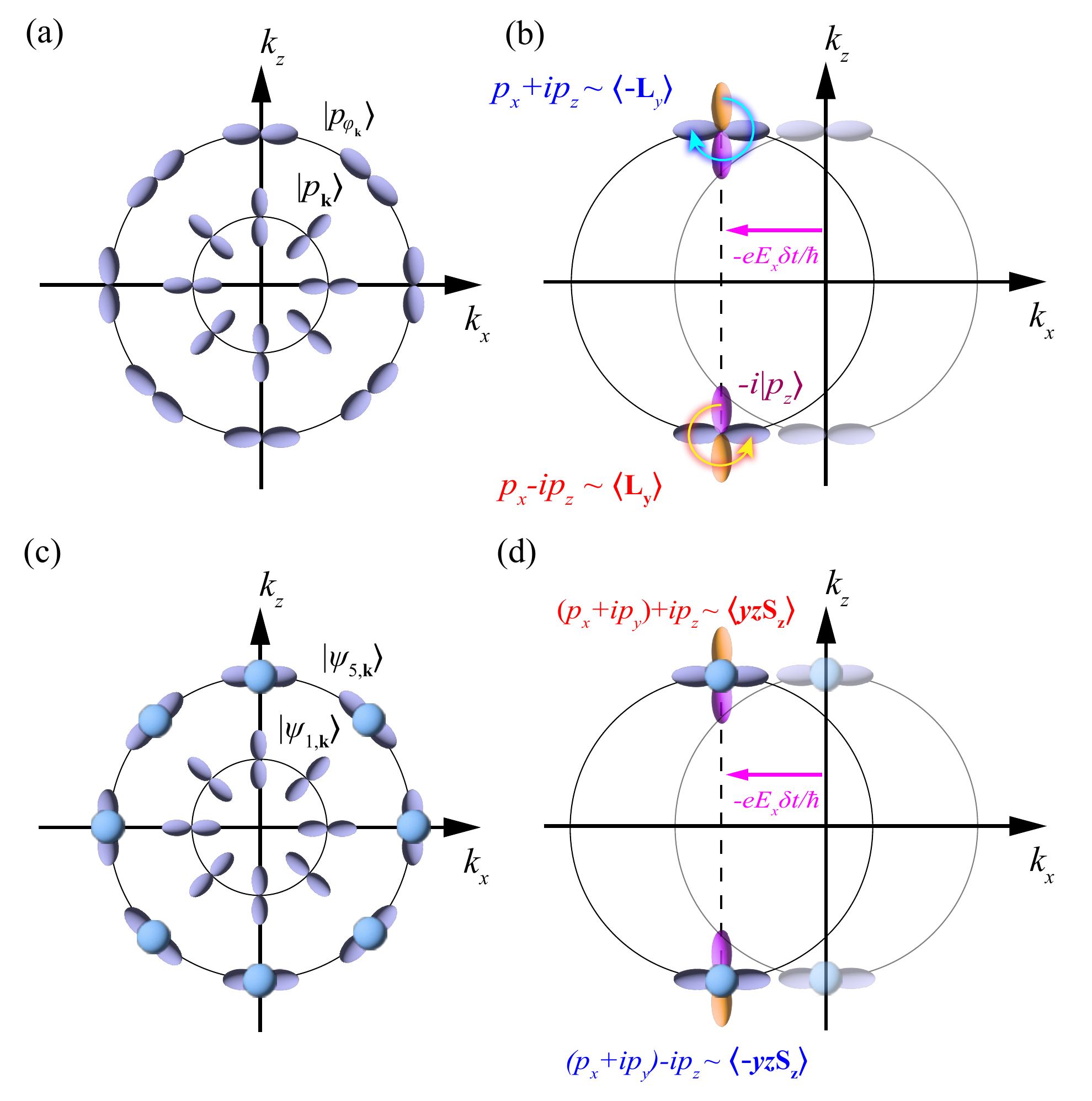}
 \caption{
(a) The eigenstates of Eq.~\eqref{eq:7} with no SOC. (b) When an electric field is applied, $\bold{k}$ shifts, leading to an imaginary mixing between the two states $\ket{p_{\phi_{\bold{k}}}}$ and $\ket{p_{\bold{k}}}$ leading to the nonvanishing OAM densities (i.e., $\ket{p_x}+i\ket{p_z}, \langle L_y \rangle\neq 0)$, whose values are opposite for $k_z > 0$ and $k_z < 0$.
(c) The eigenstates of Eq.~\eqref{eq:7} with SOC. We plot only $\ket{\psi_{1, \bold{k}}}$ and $\ket{\psi_{5, \bold{k}}}$, omitting spin for simplicity. Here, the blue sphere represents the $\ket{p_y}$ orbitals. When SOC is turned on, the $\ket{p_t}$ orbital becomes $\ket{p_t} + i\ket{p_y}$.
(d) When an electric field is applied, similar to (b), the two states $\ket{\psi_{1, \bold{k}}}$ and $\ket{\psi_{5, \bold{k}}}$ are mixed. However, there is an additional contribution due to state change (i.e., $\ket{p_x + i p_y} + i\ket{p_z}$). As a result, a non-equilibrium MO density is generated.
 }
\label{fig:6}
\end{figure}

We then examine how these eigenstates respond to an external electric field $\mathcal{E}_x$ and investigate the resulting MOHE on the $k_y=0$ plane, analogous to the previous study on OHE. We consider $\ket{\psi_{5,\bold{k}}}$ and $\ket{\psi_{6,\bold{k}}}$ as states near the Fermi energy and examine the interband mixing between $\ket{\psi_{5\bold{k}}}$ ($\ket{\psi_{5\bold{k}}}$) and $\ket{\psi_{1\bold{k}}}$, $\ket{\psi_{3\bold{k}}}$ ($\ket{\psi_{2\bold{k}}}$, $\ket{\psi_{4\bold{k}}}$) states induced by an external electric field. When $\bold{k}$ is shifted to $\bold{k}+\delta\bold{k}$, the shifted state $\ket{\tilde{\psi}_{5,\bold{k}}}$ can be decomposed into $\ket{\psi_{5,\bold{k}+\delta\bold{k}}} + \delta\phi\{(\sqrt{2}\cos\alpha + \sin\alpha)/2 \ket{\psi_{1,\bold{k}+\delta\bold{k}}} + (\sqrt{2}\sin\alpha + \cos\alpha)/2 \ket{\psi_{3,\bold{k}+\delta\bold{k}}}\}$. Then, after a short time $\delta t$, these two states acquire different phase factors $\ket{\psi_{5,\bold{k}+\delta\bold{k}}}e^{-iE_{5,\bold{k}+\delta\bold{k}}\delta t} + \delta\phi\{(\sqrt{2}\cos\alpha + \sin\alpha)/2 \ket{\psi_{1,\bold{k}+\delta\bold{k}}}e^{-iE_{1,\bold{k}+\delta\bold{k}}\delta t} + (\sqrt{2}\sin\alpha + \cos\alpha)/2 \ket{\psi_{3,\bold{k}+\delta\bold{k}}}e^{-iE_{3,\bold{k}+\delta\bold{k}}\delta t}\}$.
The nonzero expectation values of each MO component and spin component $S_y$ for the shifted state $\ket{\tilde{\psi}_{5,\bold{k}},\delta t} $ after $\delta t$ are given by
\begin{subequations}\label{eq:15}
    \begin{align}
        &\bra{\tilde{\psi}_{5,\bold{k}},\delta t}\{L_x,L_y\}S_x \ket{\tilde{\psi}_{5,\bold{k}},\delta t} \nonumber \\
        &= \delta \phi\cos^2\phi \Bigg\{ \text{Im} [e^{-i(E_{5,\bold{k}}-E_{1,\bold{k}})\delta t}]  
        \left(\cos^2\alpha+\frac{\sqrt{2}}{2}\cos\alpha\sin\alpha\right) \nonumber \\
        &\qquad \qquad \qquad - e^{-i(E_{5,\bold{k}}-E_{3,\bold{k}})\delta t}]  
        \left(\sin^2\alpha+\frac{\sqrt{2}}{2}\cos\alpha\sin\alpha\right) \Bigg\} \label{eq:15a}, \\
         &\bra{\tilde{\psi}_{5,\bold{k}},\delta t}\{L_y,L_z\}S_z \ket{\tilde{\psi}_{5,\bold{k}},\delta t}  \nonumber \\
         &= \delta \phi\sin^2\phi \Bigg\{ \text{Im} [e^{-i(E_{5,\bold{k}}-E_{1,\bold{k}})\delta t}]  
        \left(\cos^2\alpha+\frac{\sqrt{2}}{2}\cos\alpha\sin\alpha\right) \nonumber \\
         &\qquad \qquad \qquad - e^{-i(E_{5,\bold{k}}-E_{3,\bold{k}})\delta t}]  
        \left(\sin^2\alpha+\frac{\sqrt{2}}{2}\cos\alpha\sin\alpha\right) \Bigg\} \label{eq:15b}, \\
         &\bra{\tilde{\psi}_{5,\bold{k}},\delta t} S_y \ket{\tilde{\psi}_{5,\bold{k}},\delta t}  \nonumber \\
         &= \delta \phi \Bigg\{ \text{Im} [e^{-i(E_{5,\bold{k}}-E_{1,\bold{k}})\delta t}]  
        \left(\sqrt{2} \cos\alpha\sin\alpha+\sin^2\alpha\right) \nonumber \\
         &\qquad \quad - e^{-i(E_{5,\bold{k}}-E_{3,\bold{k}})\delta t}]  
        \left(\sqrt{2} \sin^2\alpha+\cos\alpha\sin\alpha\right) \Bigg\} \label{eq:15c},
    \end{align}
\end{subequations} 
Straightforward calculation reveals that $\delta\phi\propto k_z \propto \sin \phi$. Therefore, Eq.~\eqref{eq:15} shows that there exists $z$-flow of $O_{xy}^x$ ,$O_{yz}^z$, and $S_y$ which are nothing but MOHE [Fig.~\ref{fig:1}(c)] and SHE~\cite{go2018}. The other degenerate state, $\ket{\tilde{\psi}_{4,\bold{k}}}$, also contributes to the same result, and together they generate the MOHE. As a side remark, we note that each contribution of $\ket{\tilde{\psi}_{n,\bold{k}}}$ to the MOHE is nonzero even when SOC = 0 ($\alpha = 0$). However, when SOC = 0, $\ket{\psi_{3,\bold{k}}}$, $\ket{\psi_{4,\bold{k}}}$, $\ket{\psi_{5,\bold{k}}}$, and $\ket{\psi_{6,\bold{k}}}$ become degenerate ($E_{+,\bold{k}}=E_{-,\bold{k}}$), and the contributions from $\ket{\psi_{3,\bold{k}}}$ and $\ket{\psi_{4,\bold{k}}}$ are opposite to those from $\ket{\psi_{5,\bold{k}}}$ and $\ket{\psi_{6,\bold{k}}}$. Thus, their net contribution to the MOHE vanishes. For this reason, to avoid this cancellation and generate nonvanishing MOHE, the SOC is necessary.

This model also explains the similarity between the SHC and the MOHC from their angular distribution in $k_x$-$k_z$ plane. Using $\delta\phi\propto k_z \propto \sin \phi$, the $\phi$ dependencies of the SHC and the MOHC are $\sigma_{zx}^{y} \sim \sin \phi$, $\chi_{zx}^{O_{xy}^{x}} \sim \sin \phi \cos^2 \phi \sim 1/4 \sin \phi + 1/4 \sin 3 \phi$, and $\chi_{zx}^{O_{yz}^{z}} \sim \sin^3 \phi \sim 3/4 \sin \phi - 1/4 \sin 3 \phi$, respectively. Note that MOHC $\chi_{zx}^{O_{yz}^{z}}$ has $\sin\phi$ compoent three times larger than $\sin 3\phi$ component, while $\chi_{zx}^{O_{xy}^x }$ has $\sin\phi$ component equivalent to $\sin 3\phi$ component. That is, the MOHC $\chi_{zx}^{O_{yz}^{z}}$ has the angular dependency close to the SHC, while the MOHC $\chi_{zx}^{O_{xy}^x }$ has the angular dependency different from the SHC. This explains our first-principle calculation results showing that the MOHC $\chi_{zx}^{O_{yz}^{z}}$ exhibits a similar sign trend to the SHC with respect to spin-orbit coupling while the MOHC $\chi_{zx}^{O_{xy}^x }$ remains consistently negative [Fig.~\ref{fig:3}(b)].

We also provide an intuitive physical picture of MOHE. To make the explanation simple, we discuss it in the limit where $\sin\alpha\approx 0$  and $\cos\alpha\approx 1$. In this case, $\ket{\psi_{5,\bold{k}}}$ is equal to $(-\sin \phi_\bold{k}\ket{p_x} + \cos\phi_{\bold{k}}\ket{p_z} + i\ket{p_y})/\sqrt{2}$, and $\ket{\psi_{1,\bold{k}}}$ is equal to $\cos \phi_\bold{k}\ket{p_x} + \sin\phi_{\bold{k}}\ket{p_z}$ [Fig.~\ref{fig:6}(c)]. When an electric field is applied, the two states are mixed as previously described. As discussed in the two previous examples, the two states mix in an imaginary way proportional to $\text{Im}[e^{-i(E_{5,\bold{k}}-E_{1,\bold{k}})}\delta t]$. Simply, we can write this as $\left(-\sin \phi_\bold{k}\ket{p_x} + \cos\phi_{\bold{k}}\ket{p_z} + i\ket{p_y}/\sqrt{2}\right) + i\text{Im}[e^{-i(E_{5,\bold{k}}-E_{1,\bold{k}})\delta t}]\left(\cos \phi_\bold{k}\ket{p_x} + \sin\phi_{\bold{k}}\ket{p_z}\right)$. Here, $-\sin \phi_\bold{k}\ket{p_x} + \cos\phi_{\bold{k}}\ket{p_z}$ and $i (\cos \phi_\bold{k}\ket{p_x} + \sin\phi_{\bold{k}}\ket{p_z})$ combine to form the OAM $L_y$, which leads to the occurrence of OHE. Furthermore, there exist terms $i\ket{p_y}$ and $i(\cos \phi_\bold{k}\ket{p_x} + \sin\phi_{\bold{k}}\ket{p_z}$, which contribute to making the terms $\{L_x, L_y\}$ (\textit{real} mixing between $\ket{p_x}$ and $\ket{p_y}$ orbitals) and $\{L_y, L_z\}$ (\textit{real} mixing between $\ket{p_y}$ and $\ket{p_z}$ orbitals) nonzero [Fig.~\ref{fig:6} (d)]. These terms combine with spin to create the components $O_{xy}^x$ and $O_{yz}^z$, which lead to MOHE. We also perform tight-binding calculation and confirm that orbital texture and SOC are the origin of the MOHE in the Appendix B.

\section{Discussion}

We discuss the device application potential of the MO current generated by the MOHE. First, when the MO current is injected into a $d$-wave AM, it can provide torque, referred to as magnetic octupole torque (MOT)~\cite{han2024}. The MO current capable of exerting torque is determined by the coupling $\bold{N} \cdot \bold{O}_{ij}$ in the $d$-wave AM. The relevant indices $i$ and $j$ depend on the situation. For instance, in RuO$_2$[001] and MnF$_2$~\cite{bose2022, bhowal2024}, this corresponds to the $\bold{O}_{xy}$ component, while in RuO$_2$[101]~\cite{bose2022}, it involves both $\bold{O}_{xy}$ and $\bold{O}_{yz}$ components. This means that the two types of MO currents we have previously investigated, $j_z^{O_{xy}^x}$ and $j_z^{O_{yz}^z}$, can both generate torque according to the symmetry of the $d$-wave AM. The existing spin-orbit torque (SOT) can also exert torque on the $d$-wave AM. When combined with MOT, it offers richer possibilities for controlling the N\'eel vector in $d$-wave AMs. Thus, understanding the relative magnitudes of the SHC and MOHC in each material becomes important. For this purpose, we compiled a table comparing SHC (taken from Ref.~\cite{go2024}) and MOHC for various materials. For instance, if one aims to isolate the effect of MO currents—such as studying MOT-driven N\'eel vector dynamics in $d$-wave AMs—materials with small SHC but large MOHC, like hcp Zr and hcp Hf, are preferable. On the other hand, if one seeks to explore phenomena where both SOT and MOT contribute, materials with large SHC and MOHC, such as fcc Pt, fcc Rh, fcc Pd, and bcc W, are suitable candidates.
% However, the existing spin-orbit torque (SOT) can also exert torque on the $d$-wave AM, complicating the investigation of the effects induced by MOT when both effects are present. Therefore, to focus on the MOT, it is desired to identify materials where the SHC is weak and the MOHC is stronFor the comparison between the MOHC and the SHC for each metal, we list its SHC along the MOHC in Table.~\ref{tab:1}. Metals with large SHCs, such as fcc Pt, fcc Rh, fcc Pd, and bcc W, also have large MOHCs $\chi_{zx}^{O_{xy}^{x}}$ and $\chi_{zx}^{O_{yz}^{z}}$. However, there are some metals with large MOHCs but negligible SHC. For example, hcp Zr and hcp Hf have large MOHCs with small SHC: $\chi_{zx}^{O_{xy}^{x}} = -319  \ (\hbar/e) \ (\Omega \ \rm{cm})^{-1}$, $\chi_{zx}^{O_{yz}^{z}} = -607 \ (\hbar/e) \ (\Omega \ \rm{cm})^{-1}$, $\sigma_{zx}^{y} = -30 \ (\hbar/e) \ (\Omega \ \rm{cm})^{-1}$ for hcp Zr and $\chi_{zx}^{O_{xy}^{x}} = -194  \ (\hbar/e) \ (\Omega \ \rm{cm})^{-1}$, $\chi_{zx}^{O_{yz}^{z}} = -584 \ (\hbar/e) \ (\Omega \ \rm{cm})^{-1}$, $\sigma_{zx}^{y} = 50 \ (\hbar/e) \ (\Omega \ \rm{cm})^{-1}$ for hcp Hf, respectively. Those materials are suitable to generate the MOT-induced N\`eel order dynamics. 

As a second application of the MO current, we propose the MO Hall magnetoresistance, which is the MO counterpart of the spin Hall magnetoresistance~\cite{nakayama2013}. In an FM/heavy metal (HM) bilayer structure, applying an electric field changes the longitudinal resistance based on the magnetization direction of the FM, which is referred to as the spin Hall magnetoresistance. Similarly, in an AM/HM structure, we expect the resistance to vary according to the direction of the N\'eel vector, resulting in MO Hall magnetoresistance.

Finally, our work can also be expanded to higher-order multipole currents. Recently, some \textit{g}-wave AM candidates, such as $\alpha$-Fe$_2$O$_3$, CrSb, and MnTe~\cite{mcclarty2024,verbeek2024} have been reported to have magnetotriakontadipolar order. By obtaining the magnetotriakontadipole operator from the electric hexadecapole operators, one can calculate the current of magnetotriakontadipole, which could exert torque on these materials.

\section{Conclusion}

In this paper, by first-principle calculation, we show that the \textit{4d} and \textit{5d} transition metals show a large MOHC. We find that some of the transition metals have MOHCs much larger than those of their SHCs. This feature could be utilized to investigate MO physics. We also reveal the microscopic origin of MOHE by using the simple model with orbital texture and spin-orbit coupling. We also propose possible applications of the MO current generated by MOHE, including MO Hall magnetoresistance.

\begin{acknowledgments}
We thank Daegeun Jo for the fruitful discussions. I.B., S.H., and H.-W.L. were financially supported by the National Research Foundation of Korea (NRF) grant funded by the Korean government (MSIT) (No. RS-2024-00356270). S.H. was financially supported by the NRF grant funded by the Korean government (No. RS-2024-00334933). Supercomputing resources, including technical support, were provided by the Supercomputing Center, Korea Institute of Science and Technology Information (Contract No. KSC-2023-CRE-0390).
\end{acknowledgments}

\appendix

\section{Decomposition of MOHC}

\begin{figure}[t]
\includegraphics[width=8.5cm]{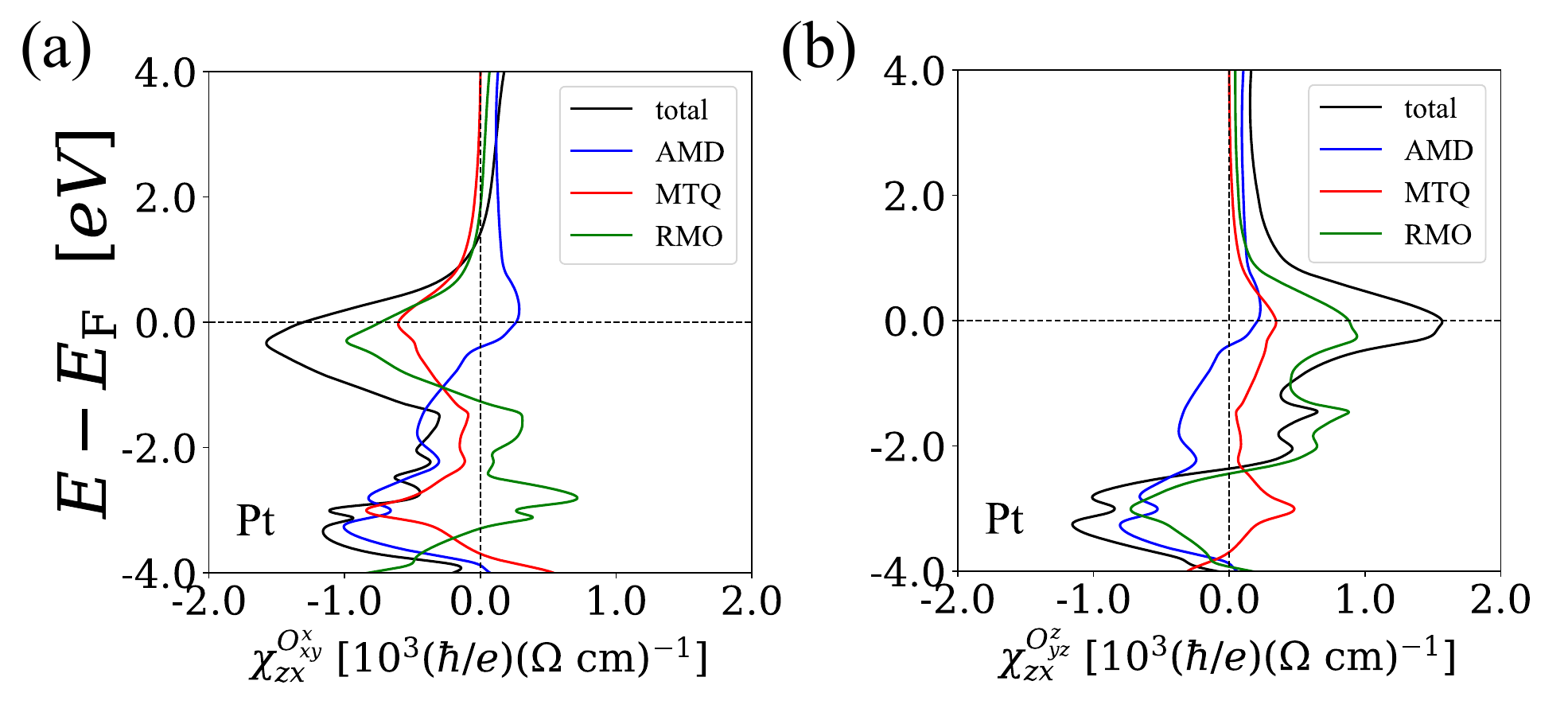}
 \caption{
The total (dark), the AMD (blue), the MTQ (red), and the RMO (green) contributions of the MOHCs $\chi_{zx}^{O_{mn}^{q}}$ for fcc Pt with MO components $O_{xy}^{x}$ (a) and $O_{yz}^{z}$ (b).
}
\label{fig:A1}
\end{figure}

We define the MO moment as a rank-3 tensor operator of the form $r_n r_m S_q$ and investigate its corresponding Hall current response. This MO operator is generally reducible and can be decomposed into rank-1 vector, rank-2 tensor, and rank-3 tensor components~\cite{hayami2020, urru2022, hayami2024}. In the following, we analyze which components contribute to the MOHC by reducing the MO tensor into these irreducible parts. The rank-1 vectors of the MO are defined as anisotropic magnetic dipoles (AMDs)~\cite{kusunose2008, kuramoto2008, hayami2021}, while the rank-2 and rank-3 tensors are called magnetic toroidal quadrupoles (MTQs) and reduced MOs (RMOs)~\cite{urru2022}, respectively. AMD is defined as $\textbf{M}^{\prime} = 3 / \sqrt{10} \{3 (\textbf{S} \cdot \textbf{r}) \textbf{r} - r^2 \textbf{S}\}$. Each component of MTQ $T_{2,m}$ with an index $m = -2, -1, 0, 1, 2$ is defined as~\cite{hayami2020}
\begin{align}
        T_{2,2} &= \frac{1}{\sqrt{2}} (yz S_x + xz S_y -2xy S_z), \nonumber\\
        T_{2,1} &= \frac{1}{\sqrt{2}} [xy S_x + (z^2 - x^2) S_y - yz S_z], \nonumber\\
        T_{2,0} &= \sqrt{\frac{3}{2}} (yz S_x - xz S_y), \nonumber\\
        T_{2,-1} &= \frac{1}{\sqrt{2}} [ (y^2 - z^2) S_x -xy S_y + xz S_z], \nonumber\\
        T_{2,-2} &= \frac{1}{\sqrt{2}} [-xz S_x + yz S_y + (x^2 - y^2) S_z]. 
\end{align}

\begin{figure}[t]
\includegraphics[width=8.5cm]{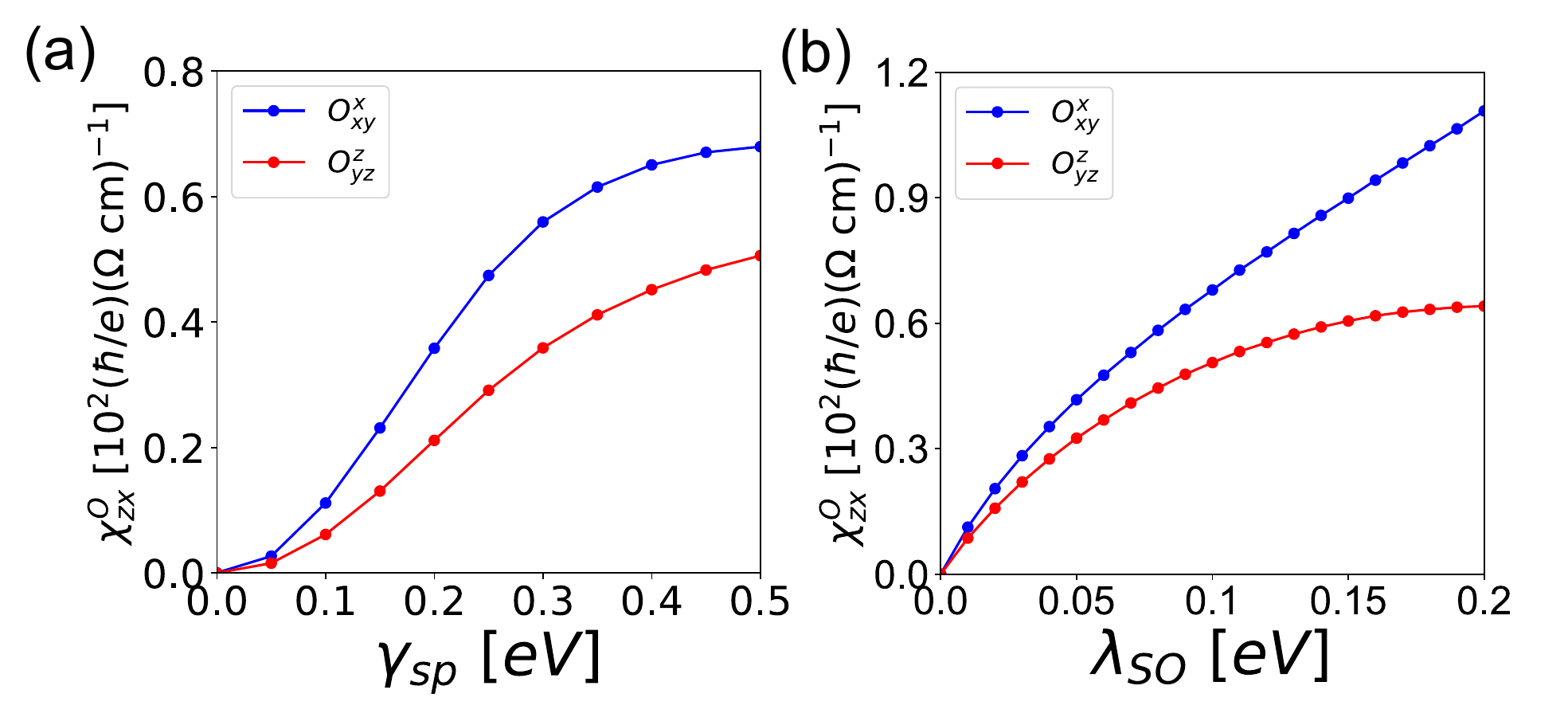}
 \caption{
The MOHCs $\chi_{zx}^{O_{mn}^{q}}$ for the tight-binding model Hamiltonian from Ref.~\cite{go2018} with MO components $O_{xy}^{x}$ (blue) and $O_{yz}^{z}$ (red), with respect to $sp$-orbital hybridization $\gamma_{sp}$ (a) and spin-orbit coupling (b).
}
\label{fig:A2}
\end{figure}

Likewise, each component of RMO $O_{3,m}$ with an index $m = -3, -2, -1, 0, 1, 2, 3$ is defined as
\begin{align}
        O_{3,3} &= \sqrt{\frac{3}{2}} [(x^2 - y^2) S_x - xy S_y ], \nonumber\\
        O_{3,2} &= xz S_x - yz S_y + \frac{1}{2} (x^2 - y^2) S_z, \nonumber\\
        O_{3,1} &= \frac{1}{\sqrt{10}} [\frac{1}{2} (4z^2 - 3x^2 - y^2) S_x - xy S_y + 4xz S_z ], \nonumber\\
        O_{3,0} &= \sqrt{\frac{3}{5}} [ -xz S_x - yz S_y + \frac{1}{2} (3z^2 - r^2) S_z ],\nonumber\\
        O_{3,-1} &= \frac{1}{\sqrt{10}} [ -xy S_x + \frac{1}{2} (4x^2 - s^2 - 3y^2) S_y + 4yz S_z], \nonumber\\
        O_{3,-2} &= yz S_x + xz S_y + xy S_z, \nonumber\\
        O_{3,-3} &= \sqrt{\frac{3}{2}} [xy S_x + \frac{1}{2} (x^2 - y^2) S_y ].
\end{align}
From the definition of AMD, MTQ, and RMO, one can decompose the MOs $xy S_x$ and $yz S_y$ are decomposed as
\begin{align}\label{eq:a3}
        xyS_x &= \frac{5}{24} \sqrt{\frac{5}{2}} M_y^{\prime} + \frac{7}{24 \sqrt{2}} T_{2,-1} + \frac{1}{\sqrt{6}} O_{3,-3} - \frac{1}{12} \sqrt{\frac{5}{2}} O_{3,-1}, \nonumber\\
        yzS_z &= \frac{1}{12} \sqrt{\frac{5}{2}} M_y^{\prime} - \frac{1}{6 \sqrt{2}} T_{2,-1} + \frac{1}{3} \sqrt{\frac{5}{2}} O_{3,-1}.
\end{align}
Using Eq.~\eqref{eq:a3}, the MOHCs $\chi_{zx}^{O_{mn}^{q}}$ with MO components $O_{xy}^{x} \sim xy S_x$ and $O_{yz}^{z} \sim yz S_z$ can also be decomposed as the AMD, MTQ, and RMO contributions.

Figure~\ref{fig:A1} shows the MOHCs $\chi_{zx}^{O_{mn}^{q}}$ for fcc Pt with MO components $O_{xy}^{x}$ (a) and $O_{yz}^{z}$ (b), plotted as total (dark) MOHCs with their AMD (blue), MTQ (red) and RMO (green) contributions. One can observe that the MO current with with MO component $O_{xy}^x$ and $O_{yz}^z$ are both decomposed into the AMD, the MTQ, and the RMO current, and that the largest contribution to the MO current (both $O_{xy}^x$ and $O_{yz}^z$) comes from the RMO current.

\section{Tight-binding analysis of the microscopic origin of MOHE}

To complement the model Hamiltonian analysis, we performed a tight-binding analysis to further investigate the microscopic origin of the MOHE. Through this analysis, we found that—similar to the spin Hall effect~\cite{go2018}—the MOHC is primarily governed by (i) the strength of the orbital texture and (ii) the magnitude of SOC.

The details of the calculation are as follows. We followed the simple cubic lattice model introduced in Ref.~\cite{go2018}, which consists of an $sp$-orbital system with nearest-neighbor hopping. Figure~\ref{fig:A2} shows the dependence of the MOHC $\chi_{zx}^O$ with the MO components $O_{xy}^x$ (blue) and $O_{yz}^z$ (red)—on (a) the orbital hybridization strength ($\gamma_{sp}$) and (b) the SOC strength ($\lambda_{\text{SO}}$). In this model, the parameter $\gamma_{sp}$ represents the strength of $sp$-orbital hybridization. This term effectively mixes the $p_x$, $p_y$, and $p_z$ orbitals via $sp$-hybridization (note that without such mixing, $p_x$, $p_y$, and $p_z$ orbitals do not directly couple through nearest-neighbor hopping), and thereby plays a crucial role in generating orbital texture~\cite{go2018,han2023}. Our results indicate that the MOHC scales proportionally with both the orbital texture strength and SOC. This proportionality provides support for the understanding that the MOHE shares the same microscopic origin as the SHE.

\bibliography{sample}

\end{document}